\documentclass[pra,twocolumn,preprintnumbers,amsmath,amssymb]{revtex4}

\usepackage{graphicx}
\usepackage{dcolumn}
\usepackage{bm}
\usepackage{color}
\usepackage{epstopdf}

\begin{document}



\title{Weak Langmuir turbulence in disordered multimode optical fibers}

\author{Kilian Baudin$^{1}$, Josselin Garnier$^{2}$, Adrien Fusaro$^{1,3}$, Nicolas Berti$^{1}$,  Guy Millot$^{1}$, Antonio Picozzi$^{1}$}
\affiliation{$^{1}$ Laboratoire Interdisciplinaire Carnot de Bourgogne, CNRS, Universit\'e Bourgogne Franche-Comt\'e, Dijon, France}
\affiliation{$^{2}$ CMAP, CNRS, Ecole Polytechnique, Institut Polytechnique de Paris, 91128 Palaiseau Cedex, France}
\affiliation{$^{3}$ CEA, DAM, DIF, F-91297 Arpajon Cedex, France} 


\begin{abstract}
We consider the propagation of temporally incoherent waves in multimode optical fibers (MMFs) in the framework of the multimode nonlinear Schr\"odinger (NLS) equation accounting for the impact of the  natural structural disorder that affects light propagation in standard MMFs (random mode coupling and polarization fluctuations).
By averaging the dynamics over the fast disordered fluctuations, we derive a Manakov equation from the multimode NLS equation, which reveals that the Raman effect introduces a previously unrecognized nonlinear coupling among the modes. 
Applying the wave turbulence theory on the Manakov equation, we derive a very simple scalar kinetic equation describing the evolution of the multimode incoherent waves.
The structure of the kinetic equation is analogous to that developed in plasma physics to describe weak Langmuir turbulence.
The extreme simplicity of the derived kinetic equation provides physical insight into the multimode incoherent wave dynamics.
It reveals the existence of different  collective behaviors where all modes self-consistently form a multimode spectral incoherent soliton state.
Such an  incoherent soliton can exhibit a discrete behavior characterized by collective synchronized spectral oscillations in frequency space. 
The theory is validated by accurate numerical simulations: The simulations of the generalized multimode NLS equation are found in quantitative agreement with those of the derived scalar kinetic equation without using adjustable parameters. 
\end{abstract}


\maketitle

\section{Introduction}

Multimode optical fibers (MMF) constitute ideal test-beds for the study of complex spatio-temporal nonlinear optical phenomena. 
The phenomena that can be tested include multi-octave spanning supercontinuum generation involving intense visible frequency combs, multiple filamentation processes, or multimode solitons \cite{wright15np,renninger13,krupa16_geom,conforti17,krupa19,agrawal}.
Actually, light dynamics in MMFs involves a variety of nonlinear effects whose complexity requires a deep understanding of spatiotemporal nonlinear propagation, with a multitude of applications such as the improvement of optical signal processing techniques for spatial division multiplexing \cite{kaminow13}, or the development of novel high-energy versatile fibre sources  \cite{wright17sc}.

Aside from potential applications, MMFs also provide a natural platform for the study of the interplay of nonlinearity and disorder \cite{leonetti14,schirmacher18,mafi19}, which is a fundamental problem of general interest \cite{silberberg08,segev,delre17,conti11,leuzzi15}.
As a matter of fact, light propagation in a conventional MMF is known to be affected by a structural disorder of the material due to inherent imperfections and external perturbations (e.g., bending, twisting,
tensions, or core-size variations in the fabrication process) \cite{kaminow13}, a feature which is relevant to endoscopic imaging for instance \cite{psaltis}.
When such a natural disorder of the fiber dominates over nonlinear effects, the nonlinear Schr\"odinger (NLS) equation describing the propagation of light can be reduced to an effective equation through the so-called Manakov approximation, a procedure originally developed for single-mode fibers \cite{wai96} and more recently extended to MMFs \cite{mecozzi12a,mecozzi12b,mumtaz13,antonelli13,antonelli16,buch19}.

From a different perspective, a fundamental phenomenon of spatial beam self-organization, termed ``beam self-cleaning", has been recently discovered in (graded-index) MMFs \cite{liu16,krupa17}.
At variance with an apparently similar phenomenon driven by the dissipative Raman effect in MMFs, known as Raman beam cleanup \cite{terry07}, this self-organization process is due to a purely conservative Kerr nonlinearity \cite{krupa17}.
While the detailed understanding of spatial beam cleaning is still debated, different works indicate that certain regimes of beam self-cleaning can be described as a natural process of optical wave thermalization to thermal equilibrium \cite{PRL19,PRA19,pod19,christodoulides19,kottos20}, a feature that has been recently demonstrated experimentally \cite{PRL20,EPL21,pourbeyram_arxiv}.
This has motivated the development of a wave turbulence formalism that takes into account  the structural disorder inherent to light propagation in MMFs. 
Following this approach, a wave turbulence kinetic equation has been derived, which revealed that the structural disorder leads to a significant acceleration of the process of thermalization and condensation, a feature that can help to understand some regimes of spatial beam-cleaning in MMFs \cite{PRA19}.

Our aim in this paper is to study the interplay of disorder and nonlinearity in the framework of a different regime of light propagation in MMFs.
At variance with the previous theoretical works describing the purely spatial dynamics\cite{PRL19,PRA19}, here we consider the spatio-temporal multimode dynamics where {\it temporally incoherent waves} propagate through the MMF.
On the basis of the wave turbulence theory \cite{zakharov92,newell01,nazarenko11,Newell_Rumpf,shrira_nazarenko13,PR14,turitsyn12,turitsyn15,musher95}, we show that the temporal multimode turbulent dynamics is dominated by the Raman effect.
More precisely, under the assumption that  the structural disorder dominates over nonlinear effects, we derive a multimode Manakov equation from the multimode NLS equation.
{\it The new Raman term in this Manakov equation unveils a previously unrecognized nontrivial coupling among the modes, which is responsible for a collective  behavior of the multimode incoherent field.}
Indeed, applying the wave turbulence theory to the multimode Manakov equation, we derive a simple scalar kinetic equation that governs the evolution of the temporal averaged spectrum of the  multimode optical field.
The kinetic equation has a form analogous to that developed in plasma physics to describe weak Langmuir turbulence in plasma \cite{musher95,ZakhPR85,zakharovjetp,montes79}.
The derived kinetic equation then greatly simplifies the multimode NLS equation and provides physical insight into the incoherent wave dynamics.
It reveals the existence of several multimode collective behaviors of the incoherent waves that propagate through the MMF. 
As a general rule, the multimode turbulent system exhibits a self-organization process, in which all modes self-consistently form a vector spectral incoherent soliton (VSIS).
This provides a generalization of the scalar (or bimodal) spectral incoherent solitons that were previously investigated by {\it always ignoring the structural disorder of the fiber} \cite{PRL08,PR14,agrawal,OL16,PRE11}. 
The reported VSIS can also exhibit a discrete behavior, which is characterized by collective synchronized spectral oscillations of the discrete soliton in frequency space. 
The numerical simulations of the generalized multimode NLS equation are found in remarkable quantitative agreement with the  derived kinetic equation {\it without using adjustable parameters}. 

From a broader perspective, we remark that Langmuir turbulence in the strongly nonlinear regime has been widely studied both theoretically and experimentally \cite{GoldmanRMP,RobinsonRMP97}, in particular in hydrodynamics \cite{Langmuir38,Craik76,Williams97,RobinsonRMP97}, or in laboratory \cite{Wong84,Vyacheslavov02} and space plasma experiments \cite{Sulzer94,Isham99}, while cavitating Langmuir turbulence has been evidenced in natural Earth's aurora driven by solar wind \cite{IshamPRL12}.
However, aside from preliminary experiments in \cite{PRE11},  experimental evidence of weak Langmuir turbulence has not been reported in the context of nonlinear optics \cite{PR14}.
In this work we show that random mode coupling in optical fibers has a stabilizing role on the dynamics of spectral incoherent solitons, {\it which makes disordered MMFs promising for the experimental study of weak Langmuir turbulence in optics}.

\section{Multimode NLS equation}

We consider the generalized NLS equation describing the propagation of the optical field in a multimode fiber with $N$ modes (i.e., $2N$ modes accounting for polarization effects) \cite{horak}.
Following the notations of Ref.\cite{horak}, the vector electric field can be  expanded into a superposition of the individual modes
${\bm E}({\bm r},z,t) = \sum_p  {\bm F}_p({\bm r})  A_p(z,t)$,
where ${\bm F}_p({\bm r})$ denotes the normalized transverse vector mode profile 
and $A_p(z,t)$ the modal envelope with $z$ the longitudinal propagation variable, ${\bm r}=(x,y)$ the vector in the transverse plane and $t$ the time variable.
The modal vector can be written ${\bm A}(z,t)=(A_p(z,t))_{p=1}^{2N}$, where the  components $A_{2j-1}$, $A_{2j}$  refer to the orthogonal linear polarization components of the $j-$th  mode. 
The field satisfies the generalized multimode NLS equation \cite{horak}:
\begin{eqnarray}
\label{eq:nlsmulti}
i \partial_z {\bm A} + {\bf D}(z) {\bm A} +{\bf D}_0 {\bm A} + i {\bf V} \partial_t {\bm A}
- {\bf W} \partial_{tt} {\bm A} \quad \quad \quad \nonumber \\
\quad \quad +  \gamma (1-f_R) {\bm P}({\bm A}) +  \gamma f_R  {\bm Q}({\bm A})  =0.
\label{eq:A}
\end{eqnarray}
Here, ${\bf D}_0$, ${\bf V}$ and ${\bf W}$ are deterministic $2N \times 2N$ diagonal matrices that model respectively the propagation constants, the modal inverse group velocity and the modal dispersion (relative to the fundamental fiber mode).
The terms ${\bm P}({\bm A})$ and ${\bm Q}({\bm A})$ are, respectively, the  Kerr nonlinearity and  the Raman nonlinearity, which have the general forms
\begin{align}
\big[{\bm P} ({\bm A})\big]_p &= \sum_{l,m,n=1}^{2N} S_{plmn}^K A_l A_m A_n^* ,
\label{eq:Kerr_term}\\
\big[{\bm Q}({\bm A})\big]_p &= \sum_{l,m,n=1}^{2N} S_{plmn}^R A_l  [ R \star (A_m  A_n^*)] ,
\label{eq:Raman_term}
\end{align}
where $R$ is the Raman response function, $*$ stands for complex conjugation and $\star$ denotes the convolution product.
The Raman term contributes with a fraction $f_R$ to the overall nonlinearity ($f_R = 0.18$ for silica glass fibres) \cite{agrawal}.
The nonlinear coefficient is $\gamma=n_2 \omega_0/c=n_2 k_0$, where $\omega_0$ is the laser carrier frequency and $\lambda=2\pi/k_0$ the corresponding wavelength.
The tensors $S_{plmn}^K$ and $S_{plmn}^R$ are given in explicit form in Ref.\cite{horak}.

We consider the regime of strong random coupling among the spatial modes and polarization states, which is relevant for large propagation lengths in the MMF, typically larger than a few hundred meters \cite{ho14}.
The most general form of random mode coupling that conserves the total power ${\cal P}=\sum_{p=1}^{2N} |A_p|^2$ is provided by a $2N \times 2N$ random matrix-valued process ${\bf D}(z)$ that is Hermitian.
Note that the structural disorder of the MMF may also affect the group-velocity and the group-velocity dispersion of the propagating modes, which can be modelled by considering random matrices ${\bf V}(z)$ and ${\bf W}(z)$ in Eq.~(\ref{eq:KE_n_A}) as will be discussed later.

\section{Manakov reduction}

We consider the so-called Manakov regime where the impact of strong linear random coupling dominates over nonlinear effects \cite{mumtaz13}, i.e., $L_{nl}\gg \ell_c, 2\pi/\sigma$, where $L_{nl}=1/(\gamma {\cal P}) $ is the nonlinear length, while $\ell_c$ and $\sigma$ denote the correlation length and standard deviation of the random process ${\bf D}(z)$ that models the structural disorder.
We recall that the Manakov reduction has already been applied to the multimode NLS equation without the Raman effect \cite{mecozzi12a,mecozzi12b,mumtaz13,buch19}. 
On the other hand, the Manakov approximation has been considered to study the Raman amplification process in Ref.\cite{antonelli13}, and applied to the multimode NLS equation accounting for the first-order correction of the Raman response \cite{antonelli16}.
Here, we generalize the derivation of the Manakov equation, which reveals that the Raman effect introduces a nontrivial coupling among the modes that plays a key role for the incoherent propagation regime.

Let us introduce the unitary matrix ${\bf U}(z)$ solution of 
$i \partial_z {\bf U} ={\bf D} {\bf U}$, with ${\bf U}(0)={\bf I}$,
and define the mode amplitudes in the local disordered axes
${\bm B} (z,t) = {\bf U}^{-1} (z) {\bm A} (z,t)$,
with ${\bf U}^{-1} (z) ={\bf U}^\dag(z)$, where the superscript $\dag$ denotes the conjugate transpose.
We now follow the idea originally introduced for single mode fibers by Wai and Menyuk \cite{wai96}, in which birefringence fluctuations are assumed so strong that the probability density of the polarization state uniformly covers the surface of the Poincar\'e sphere, so that one can average the propagation equation over all polarization states. 
By generalizing to MMFs, we assume that the linear coupling among the modes due to ${\bf D}(z)$ is the dominant effect, so that the random matrix-valued process ${\bf U}(z)$ becomes uniformly distributed in the set of unitary matrices.
In this way, we derive in the Appendix the following homogenized Manakov multimode NLS equation:
\begin{eqnarray}
i \partial_z {\bm B} +{\widetilde d} {\bm B}   +  \frac{i}{\widetilde{v}}  \partial_t {\bm B}
- \frac{\widetilde{\beta} }{2} \partial_{tt} {\bm B} \quad \quad \quad \quad \quad \quad \quad \quad \quad \nonumber \\
\quad \quad \quad \quad \quad \quad  +  \gamma (1-f_R) \widetilde{\bm P}({\bm B})
+  \gamma f_R  \widetilde{\bm Q}({\bm B})  =0  ,
\label{eq:manakovB}
\end{eqnarray}
with
\begin{align}
\big[ \widetilde{\bm P}({\bm B}) \big]_p =&
\big(  \widetilde{S}_{(1)}^K +  \widetilde{S}_{(2)}^K  \big)\Big[\sum_{l=1}^{2N} |B_l|^2 \Big]B_p, 
\label{eq:kerrman2} \\
\big[ \widetilde{\bm Q}({\bm B}) \big]_p  =&
 \widetilde{S}_{(1)}^R  \sum_{l=1}^{2N} B_l \big[ R \star( B_p B_l^*)\big] 
  \nonumber \\
& +\widetilde{S}_{(2)}^R  B_p \Big[ R \star \Big(\sum_{l=1}^{2N}  | B_l|^2\Big)\Big] ,
\label{eq:ramanman2}
\end{align}
where we have for $X\in \{K,R\}$
\begin{eqnarray*}
\widetilde{S}_{(1)}^{X} =    \frac{1}{4N^2-1} \sum_{p',l'} S_{p'l'p'l'}^{X}- \frac{1}{2N(4N^2-1)}
\sum_{p',l'} S_{p'p'l'l'}^{X} ,
\\
\widetilde{S}_{(2)}^{X}  =\frac{1}{4N^2-1} \sum_{p',l'} S_{p'p'l'l'}^{X}- \frac{1}{2N(4N^2-1)}
\sum_{p',l'} S_{p'l'p'l'}^{X}  .
\end{eqnarray*}
As already discussed in previous works \cite{mecozzi12a,mecozzi12b,mumtaz13}, Eq. (\ref{eq:kerrman2}) shows that any  instantaneous cubic nonlinearity gives rise to an effective (phase-insensitive) deterministic Manakov-type nonlinear term.
Eq.~(\ref{eq:ramanman2}) also shows the contribution of the Raman effect, which gives rise to an effective Raman-type nonlinear term that depends only on two parameters, $ \widetilde{S}_{(1)}^R$ and $ \widetilde{S}_{(2)}^R$. 

Note that, as a result of the Manakov averaging procedure, the propagation constant, the group velocity and the dispersion coefficients have been homogenized in (\ref{eq:manakovB}):
\begin{equation*}
\widetilde{d} = \frac{1}{2N} {\rm Tr}({\bf D}_0), \quad   
\frac{1}{\widetilde{v}} = \frac{1}{2N} {\rm Tr}({\bf V}), \quad   
\widetilde{\beta} = \frac{1}{N} {\rm Tr}({\bf W}).
\end{equation*}
Accordingly, all of the modes evolve with the same propagation constant, group velocity and  dispersion coefficient.
Note however that this results from the assumption that the matrices ${\bf V}$ and ${\bf W}$ in (\ref{eq:nlsmulti}) are deterministic and constant.
If the matrix ${\bf V}(z)$ in (\ref{eq:nlsmulti}) was randomly varying, with random fluctuations that could be correlated to those of ${\bf D}(z)$, then the transport term in (\ref{eq:manakovB}) would have the form $\widetilde{\bf V}  \partial_t {\bm B}$ instead of $\frac{1}{\widetilde{v}}  \partial_t {\bm B}$, with the effective matrix 
$\widetilde{\bf V} = \left< {\bf U}^\dag(z) {\bf V}(z) {\bf U}(z) \right> $, where $\left< \cdot \right>$ stands for the average with respect to the stationary distribution of $({\bf U}(z),{\bf V}(z))$.
The same argument holds for the dispersion effects ${\bf W}$, and the corresponding effective matrix
$\widetilde{\bf W} = \left<  {\bf U}^\dag(z) {\bf W}(z) {\bf U}(z) \right>$. 


We finally remark that the nonlinear Raman-type terms in Eq.~(\ref{eq:manakovB}) are different from those reported in \cite{antonelli16}, because in this latter work the authors made use of the assumption that the nonlinear terms are co-polarized with the field, which is not justified in general. 
In particular, when one considers the propagation of incoherent waves, {\it the terms that are not co-polarized (i.e., the ones associated with $ \widetilde{S}_{(1)}^R$ in Eq.(\ref{eq:ramanman2})) are the only ones  that give rise to a coupling among the modes}.
These terms should not be neglected in our framework, a feature that will become apparent from the weak Langmuir turbulence kinetic equation discussed in the next section.


\section{Weak Langmuir turbulence kinetic equation}

In the following we derive the weak Langmuir turbulence kinetic equation governing the evolution of the averaged spectra of the incoherent waves that propagate in the MMF.
For this purpose, we consider the weakly nonlinear regime where linear dispersion effects dominate over nonlinear effects $L_{lin,j} \ll L_{nl}$, where $L_{lin,1}=t_c \widetilde{v}$ and $L_{lin,2}=2 t_c^2/\widetilde{\beta}$ denote the characteristic propagation lengths associated to the first- and second-order dispersion effects in the modal NLS Eq.~(\ref{eq:A}), and $t_c$ is the correlation time of the incoherent waves.

We are interested in the propagation of incoherent waves, where $A_p(z=0,t)$ are random functions with fluctuations that are statistically stationary in time.
Then the components $B_p(z=0,t)$ are also random functions with statistically stationary fluctuations. 
By taking an average over the random initial conditions $B_p(z=0,t)$, we can derive the wave turbulence Langmuir kinetic equation by following the  procedure of Ref.\cite{PR14}.
Next, taking the Fourier transform, the spectra
$n_{B_p} ( \omega, t,z) =\int \left< B_p(z,t+\tau/2) B_p^*(z,t-\tau/2)\right> \exp(-i\omega \tau) d\tau$,
satisfy the multimode weak Langmuir turbulence kinetic equations:
\begin{eqnarray}
\partial_z n_{B_p}(\omega,z)=   \gamma_1 n_{B_p}(\omega) \sum_{j=1}^{2N} \int g(\omega-u) n_{B_j}(u) du 
\quad \quad  \nonumber \\
+\gamma_2 n_{B_p}(\omega) \int g(\omega-u) n_{B_p}(u) du , \quad
\label{eq:langmuirmanakov}
\end{eqnarray}
where $\gamma_j=\gamma f_R {\widetilde S}^R_{(j)}/\pi$ for $j=1,2$, and $g(\omega)=\Im[{\tilde R}(\omega)]$ denotes the Raman gain function, ${\tilde R}(\omega)=\int_0^\infty R(t) \exp(-i\omega t) dt$ being the Fourier transform of the response function. 
Note that the Raman gain $g(\omega)$ is an odd function (see the inset of Fig.~1(a)), reflecting the fact that the low-frequency components are amplified to the detriment of the high-frequency components.
We have omitted to write the time dependence for the evolution of the spectrum in Eq.(\ref{eq:langmuirmanakov}), i.e.,  $n_{B_p}(\omega,t,z) \to n_{B_p}(\omega,z)$.
Indeed, since the initial condition exhibits a stationary statistics and the kinetic Eq.(\ref{eq:langmuirmanakov}) does not explicitly involve the time variable $t$, then the stationary statistics is preserved during the propagation and the averaged spectrum does not depend on time.

It becomes apparent in the kinetic Eq.(\ref{eq:langmuirmanakov}) that only the first term proportional to $\gamma_1=\gamma f_R {\widetilde S}^R_{(1)}/\pi$ gives rise to a nonlinear coupling among the modes.
In other terms, {\it only the new terms in the derived Manakov equation  (i.e., the ones associated with $ \widetilde{S}_{(1)}^R$ in Eq.(\ref{eq:ramanman2})) give rise to a mode coupling in the kinetic Eq.(\ref{eq:langmuirmanakov}).}

It is important to note that the initial conditions in the basis  $B_p$ can be homogenized.
Indeed, let the field propagates over few correlation lengths $\ell_c$, in such a way that the matrix ${\bf U}$ becomes uniformly distributed while other linear and nonlinear effects are still negligible.
Under such circumstances, the initial conditions for $B_p(z,t)$ are such that 
$\left< B_p(z=0^+,t) B_{l}^*(z=0^+,t') \right>  = \delta_{pl}\frac{1}{2N}
\sum_{j=1}^{2N} \left< A_j(z=0,t) A_{j}^*(z=0,t') \right>$,
where $\delta_{pl}=1$ if $p=l$ and $0$ otherwise.
Taking the Fourier transform, we obtain a homogeneous initial condition for the spectra $n_{B_p}(\omega,z=0^+)=\frac{1}{2N} \sum_j n_{A_j}(\omega,z=0)$ for $p =1,\ldots,2N$.
An other important point to stress is that this homogeneous initial condition is preserved during the propagation: 
As a consequence of the averaging Manakov procedure, the modal coupling coefficients $\gamma_1$ and $\gamma_2$ in the kinetic Eqs.(\ref{eq:langmuirmanakov}) are identical for all the modes, so that the spectra $n_{B_p}(z,\omega)$  verify 
$$
n_{B_p}(\omega,z) = n_B(\omega,z) \quad {\rm for} \quad p =1,\ldots,2N.
$$
The multimode kinetic Eqs.~(\ref{eq:langmuirmanakov}) then reduce to a {\it single scalar kinetic equation}:
\begin{eqnarray}
\partial_z n_{B}(\omega,z) = (2N \gamma_1 +\gamma_2) n_{B}(\omega) \int g(\omega-u) n_{B}(u) du, 
\end{eqnarray}
with $n_{B}(\omega,z=0^+)=\frac{1}{2N} \sum_j n_{A_j}(\omega,z=0)$.

Following a similar argument, the spectra $n_{A_p} (z , \omega) =\int \left< A_p(z,t+\tau/2) A_p^*(z,t-\tau/2)\right> e^{-i\omega \tau} d\tau$ verify 
$n_{A_p}(z,\omega)=\frac{1}{2N} \sum_{j=1}^{2N} n_{B_j}(z,\omega) = n_B(\omega,z)$ for any $p$ and for $z\gg \ell_c $, even if the initial modal spectra $n_{A_p}(z=0,\omega)$ are different from each other, see Appendix.
Then we arrive at the main conclusion that the averaged spectra in the original basis ${\bm A}$ verify the {\it scalar weak Langmuir turbulence kinetic equation:}
\begin{eqnarray}
\partial_z n_{A}(\omega,z) = (2N \gamma_1 +\gamma_2) n_{A}(\omega) \int g(\omega-u) n_{A}(u) du.
\label{eq:KE_n_A}
\end{eqnarray}
We stress the remarkable simplicity of the kinetic Eq.~(\ref{eq:KE_n_A}) as compared to the original multimode NLS Eq.~(\ref{eq:A}). 
First, as in the usual scalar case, both effects of linear dispersion and instantaneous Kerr nonlinearity do not enter the kinetic equation, a property that has been confirmed by
several previous works in different circumstances \cite{PR14}.
Secondly, the structural disorder leads to an effective homogeneization that is characterized by an equipartition of the power among the modes, which further simplifies the vector kinetic Eq.~(\ref{eq:langmuirmanakov}) to the scalar kinetic Eq.~(\ref{eq:KE_n_A}).
The kinetic Eq.~(\ref{eq:KE_n_A}) has two conserved quantities, the power $P=\int n_{A}(z,\omega) d\omega$, and the `entropy' $S=\int \log[n_{A}(z,\omega)] d\omega$ \cite{PR14}.

Finally, we recall the formal analogy between the universal form of the kinetic equation describing the weakly nonlinear regime of Langmuir turbulence \cite{musher95} and the kinetic equations derived in this work.
The formal mathematical similarity mainly relies on the analogy between the molecular vibrations mediated by the optical Raman effect in optical fibers and the excitations of ion-sound waves mediated by the decay of plasma oscillations \cite{PR14}.

\section{Numerical simulations}

We have tested the validity of the theory by performing numerical simulations of the original NLS Eq.~(\ref{eq:A}) and of the derived scalar weak Langmuir turbulence kinetic Eq.~(\ref{eq:KE_n_A}).
We have considered a step-index bimodal fiber (core diameter 6$\mu$m, index difference $\Delta=0.005$, wavelength $\lambda=1.55\mu$m) in which the fundamental LP01 mode is coupled to two degenerate modes LP01a and LP01b, which results in a total of $2N=6$ coupled equations for the multimode NLS Eq.~(\ref{eq:A}).
As described in the theory, we have considered the regime of strong random linear coupling among all modes, with variance $\sigma^2$ and correlation length $\ell_c$.
We have also considered the standard form of the damped harmonic oscillator Raman function in silica optical fibers, $R(t)=H(t)(\tau_1^{-2}+\tau_2^{-2}) \tau_1 \exp(-t/\tau_2) \sin(t/\tau_1)$ with $\tau_1=12.2$fs and $\tau_2=32$fs, $H(t)$ being the Heaviside function ($\nu_R = 1/(2\pi \tau_1) \simeq 13$THz denoting the resonant Raman frequency) \cite{agrawal}.
The corresponding Raman gain function then reads 
\begin{equation}
g(\omega) = \frac{1+\eta^2}{2\eta } \Big( \frac{1}{1+(\eta+\tau_2 \omega)^2}- \frac{1}{1+(\eta-\tau_2 \omega)^2} \Big),
\end{equation}
with the time ratio $\eta=\tau_2/\tau_1$.

We consider a partially coherent optical field that is injected in the MMF and populates different modes. 
In the following, for simplicity, we assume that the different modes are initially populated with partially coherent waves with a Gaussian spectrum and random spectral phases, i.e., $A_p(z=0,t)$ has  stationary Gaussian statistics with mean zero and Gaussian covariance function and the modes $p=1,\ldots,2N$ are independent from each other.
Note that, because of the strong random coupling regime, this latter assumption is verified in practice after a propagation length $z \gg \ell_c$ that remains smaller than the nonlinear length $z < L_{nl}$.
This latter property has been verified by numerical simulations of the generalized NLS Eq.~(\ref{eq:A}).

\subsection{Multimode discrete spectral incoherent soliton}

In the following we illustrate different turbulent regimes of the system that depend on the spectral widths of the launched optical field.
Figure~1 reports a typical example of the evolution for a spectral width of the initial condition of the order of $\Delta \nu \simeq 15$THz.
Since this spectral width is of the same order as the spectral width of the Raman gain, the red-shift of the wave spectrum exhibits a discrete behavior, because the leading edge of the low-frequency tail of the spectrum exhibits a larger gain as compared to the mean gain of the whole front of the spectrum (see the inset in Fig.~1(a)).
As a result of cascaded Raman scattering \cite{agrawal,pourbeyram13}, the spectrum exhibits a discrete spectral shift that is determined by the Raman frequency $\nu_R$.
The remarkable result is that the global spectral red-shift of the field is regular and exhibits a discrete soliton-like behavior.
As already discussed in the literature \cite{PRA11,PRE11}, the discrete soliton propagates with a constant velocity in frequency space for arbitrary long distances, without emitting apparent radiation.
We recall that the spectral incoherent soliton is `hidden' in frequency space, in the sense that the soliton behavior cannot be identified in the temporal domain, where the field ${\bm A}(t,z)$ is a random wave featured by a stationary statistics \cite{PRL08}.
In this respect, the VSISs are fundamentally different in nature from optical  solitons recently investigated in MMFs \cite{Ramachandran19,agrawal19}.
Also note that a constant noise background has been added in the simulations.
Such a spectral noise is important in order to sustain a steady incoherent soliton propagation \cite{PR14}, otherwise the  soliton would undergo a slow adiabatic reshaping so as to adapt its shape to the local value of the noise background. 
This noise background can also simulate the presence of a quantum noise background.

We stress the remarkable quantitative agreement that has been obtained between the simulation of the multimode NLS Eq.(\ref{eq:nlsmulti}) and of the scalar kinetic Eq.(\ref{eq:KE_n_A}), without using any adjustable parameter.
Such a good agreement is clearly visible in the normal and logarithmic plots reported in Figs.~1(c)-(d) at a particular propagation length.
In this simulation, the different modes are initially populated with different amount of powers, as illustrated in Fig.~1(c) (gray solid lines).
As expected from the theory, we can observe in Fig.~1(e) that  random mode coupling  leads to an equipartition of power among the modes, after a propagation length of the order of the correlation length $z \gtrsim \ell_c = 10$cm.


\begin{center}
\begin{figure}
\includegraphics[width=1\columnwidth]{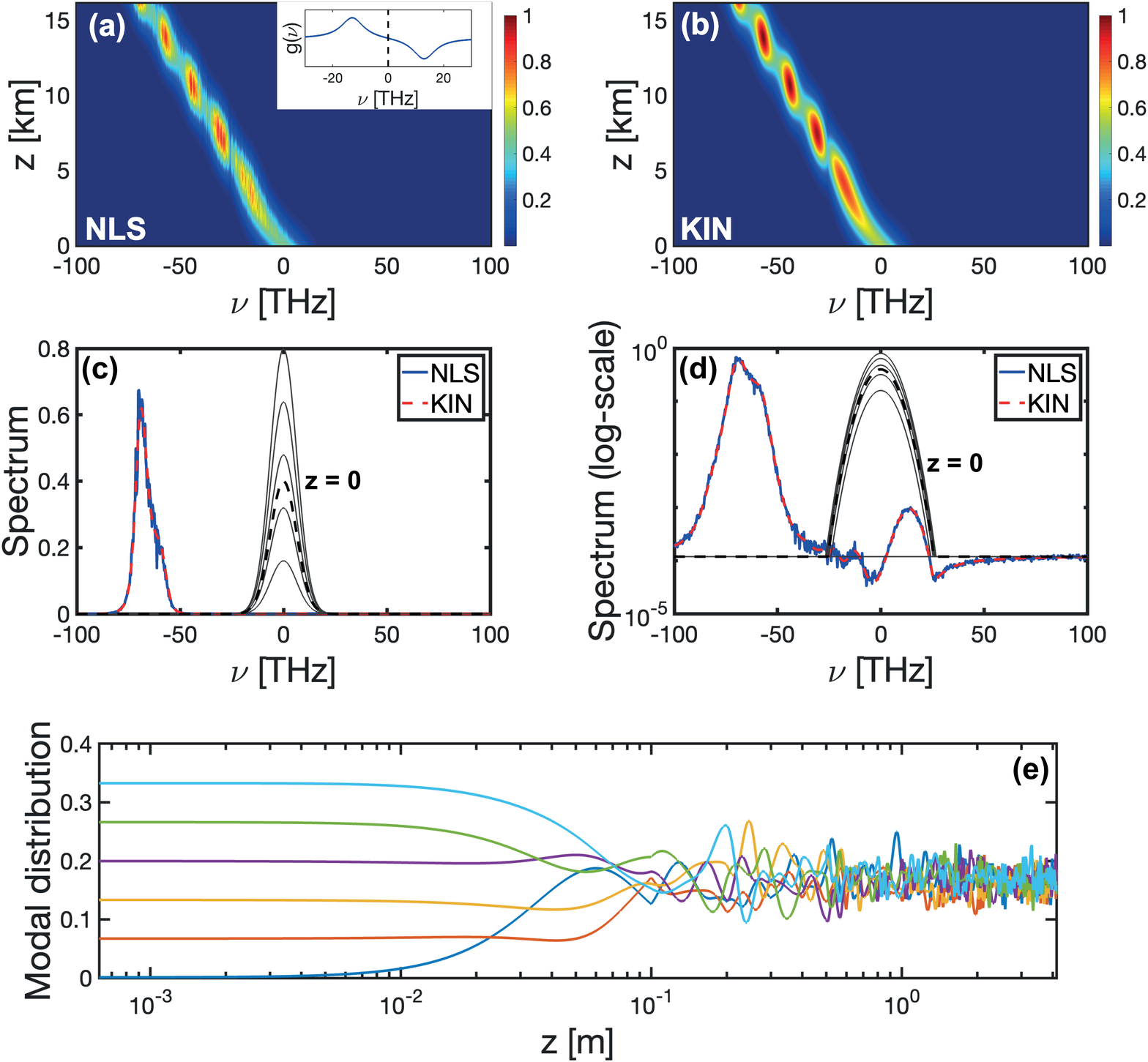}
\caption{
\baselineskip 10pt
{\bf Multimode discrete spectral incoherent soliton:}  
Evolution of the spectrum of the field during the propagation obtained by simulation of the generalized multimode NLS Eq.~(\ref{eq:A}) (a), and of the scalar weak Langmuir turbulence kinetic Eq.~(\ref{eq:KE_n_A}) (b).
Spectral profiles in normal scale (c), and logarithmic scale (d), at the propagation length $z=16$km: The blue line reports the result of the simulation of the NLS Eq.~(\ref{eq:A}) (averaged over the 6 fiber modes), the red line reports the result of the simulation of the weak Langmuir turbulence kinetic Eq.~(\ref{eq:KE_n_A}). 
The gray solid lines in (c)-(d) report the initial conditions of the six modes, while the dashed black lines the corresponding average.
(e) Evolution during the propagation ($z$ is in log-scale) of the relative amount of power of the 6 fiber modes from the simulation of the generalized NLS Eq.~(\ref{eq:A}): Random mode coupling induces an equipartition of power among the $2N=6$ modes after a propagation length $z \gtrsim \ell_c$ ($\ell_c=10$cm, $\sigma=63$m$^{-1}$, $P=17$W).
The inset in (a) shows the Raman gain spectrum $g(\nu)$, with $\nu=\omega/(2\pi)$.
The quantitative agreement between the NLS (Eq.~(\ref{eq:A})) and kinetic (Eq.~(\ref{eq:KE_n_A})) simulations is obtained without adjustable parameters.
}
\label{fig:1} 
\end{figure}
\end{center}


\subsection{Multimode continuous spectral incoherent soliton}

In this sub-section we illustrate a turbulent regime characterized by the emergence of a continuous spectral incoherent soliton.
Indeed, when the spectral width of the initial field becomes larger than the resonant Raman frequency, then the low-frequency tail of the spectrum sees a gain comparable to the mean gain of the spectral front as a whole.
In the example of Fig.~2 we have considered a spectral width $\Delta \nu \simeq 50$THz, which is much larger than in Fig.~1.
As a consequence, we can see in Fig.~2(a)-(b) that the red-shift of the wave spectrum is no longer discrete, but continuous, then giving rise to a continuous VSIS behavior.
We remark that, for the broad spectral widths considered in Fig.~2, higher-order terms should be included in the NLS model to accurately describe light propagation in the fiber \cite{horak} (also see \cite{OL14}).
However, our purpose here is just to provide a qualitative overview of different possible incoherent dynamics, while a more realistic regime of light propagation in MMFs will be considered in the next sub-section.

The continuous spectral incoherent soliton reported in Fig.~2 can be described theoretically as a stationary soliton solution of the Langmuir kinetic Eq.~(\ref{eq:KE_n_A}) \cite{PRA10}:
\begin{eqnarray}
n_{A}^{sol}({\tilde \omega}) = n_A^0 + (n_{A}^m - n_A^0) \exp\Big[-\log\big(\frac{n_A^m}{n_A^0} \big)  \frac{{\tilde \omega}^2}{\omega_0^2} \Big]  ,
\label{eq:analyt_SIS}
\end{eqnarray}
where ${\tilde \omega}=\omega - V z$, $n_A^0$ refers to the constant background noise, $n_A^m (\gg n_A^0)$ is the soliton spectral amplitude, and $\omega_0$ denotes the typical spectral width of Raman gain defined by  $\omega_{\rm 0}  = \sqrt{2} [ -\partial_\omega g(0)]^{-1/2}[ - \int_0^{\infty} g(\omega) d\omega]^{1/2}$. 
The soliton (\ref{eq:analyt_SIS}) propagates in frequency space with a constant velocity given by 
\begin{eqnarray}
V = -(2N \gamma_1+\gamma_2) 
\frac{ \int n_{A}(\omega) - n_A^0 d\omega }{\int \log(n_{A}(\omega)/n_A^0 ) d\omega}
\int \omega g(\omega) d\omega.
\label{eq:SIS_V}
\end{eqnarray}
We can observe a remarkable agreement between the analytical soliton solution given by Eq.(\ref{eq:analyt_SIS}) and the numerical simulations of both the generalized NLS Eq.(\ref{eq:nlsmulti}) and the kinetic Eq.(\ref{eq:KE_n_A}), as illustrated in Fig.~2(c)-(d).
Note however in the logarithmic plot in Fig.~2(d) a discrepancy between the solution (\ref{eq:analyt_SIS}) and the simulations in the tails of the soliton, a feature that can be explained by the fact that (\ref{eq:analyt_SIS}) is valid in the vicinity of the soliton peak.
As a matter of fact, the computation of the soliton velocity is very sensitive to the tails of the soliton profile, as revealed by the expression of $V$ in Eq.(\ref{eq:SIS_V}), whose denominator involves the logarithm of the soliton profile.
Consequently, the computation of $V$ with the analytical solution (\ref{eq:analyt_SIS}) matches the numerics qualitatively but not quantitatively, while a very good agreement of the soliton velocity (\ref{eq:SIS_V}) with the numerics is obtained by considering the soliton profile generated in the simulation, as illustrated by the dashed white lines in Fig.~2(a)-(b) that are parallel to the soliton trajectory.

\begin{center}
\begin{figure}
\includegraphics[width=1\columnwidth]{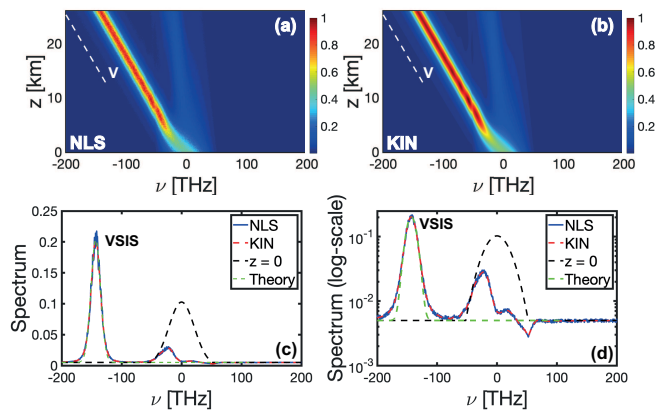}
\caption{
\baselineskip 10pt
{\bf Multimode continuous spectral incoherent soliton:}  
Evolution of the spectrum of the field during the propagation obtained by simulation of the generalized NLS Eq.~(\ref{eq:A}) (a), and of the derived scalar weak Langmuir turbulence kinetic Eq.~(\ref{eq:KE_n_A}) (b).
Spectral profiles in normal scale (c), and logarithmic scale (d), at the propagation length $z=26$km: The blue line reports the result of the NLS Eq.~(\ref{eq:A}) simulation (averaged over the 6 modes), the red line reports the result of the weak Langmuir turbulence kinetic Eq.~(\ref{eq:KE_n_A}) simulation. 
The dashed black-line reports the initial condition.
Parameters are the same as in Fig.~1 ($\ell_c=10$cm, $\sigma=63$m$^{-1}$, $P=17$W).
The spectral width is larger than in Fig.~1, which induces a continuous motion of the VSIS.
The dashed green lines in (c) and (d) report the analytical soliton solution from Eq.(\ref{eq:analyt_SIS}).
The dashed white lines in (a) and (b) denote the soliton velocity $V$ from Eq.(\ref{eq:SIS_V}) (see the text for details).
The quantitative agreement between the NLS (Eq.~(\ref{eq:A})) and kinetic (Eq.~(\ref{eq:KE_n_A})) simulations is obtained without adjustable parameters.
}
\label{fig:2} 
\end{figure}
\end{center}

\begin{center}
\begin{figure}
\includegraphics[width=1\columnwidth]{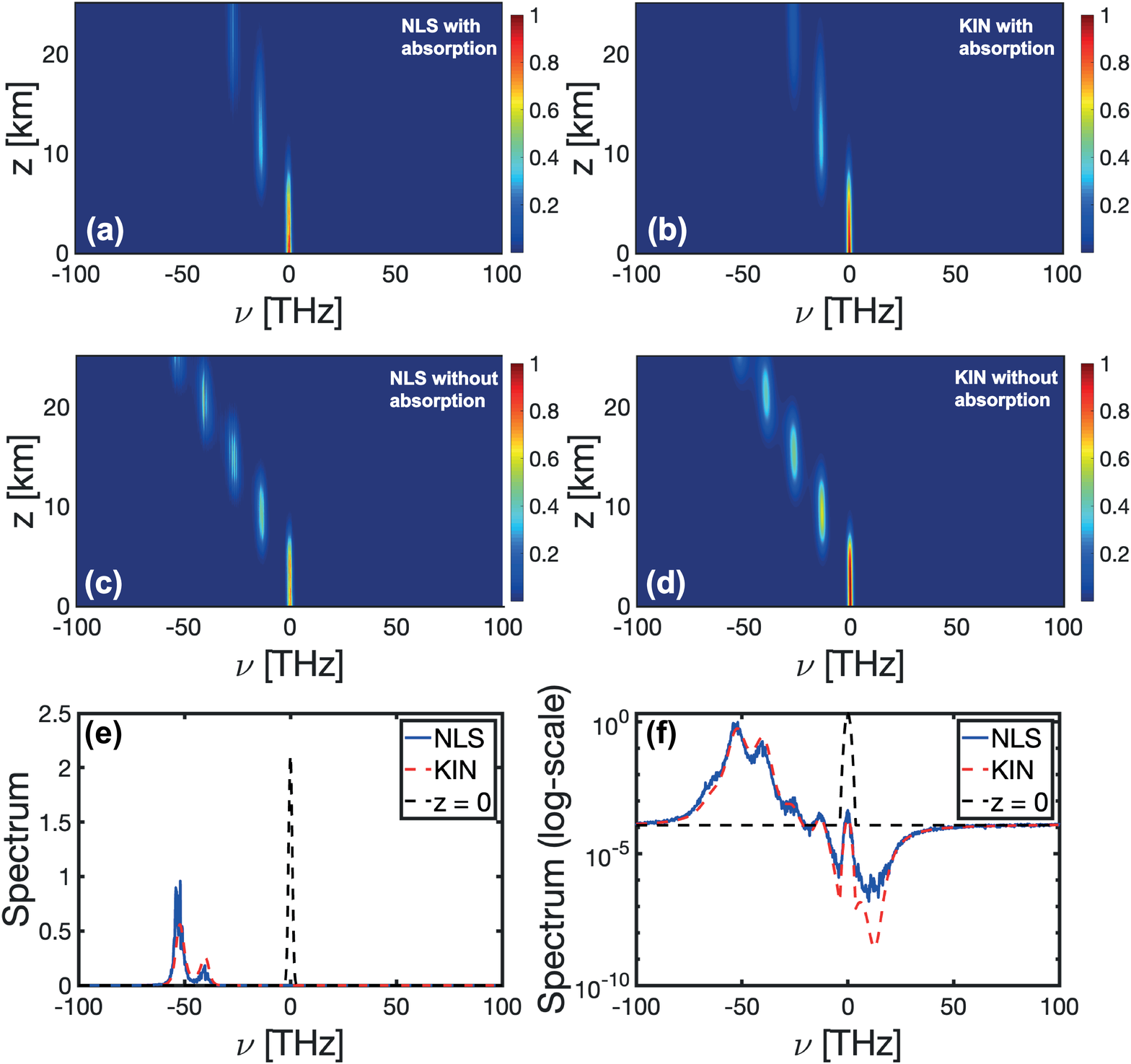}
\caption{
\baselineskip 10pt
{\bf Synchronization of incoherent spectral oscillations:}  
Evolution of the spectrum of the field during the propagation obtained by simulation of the generalized NLS Eq.~(\ref{eq:A}) (a), and of the derived scalar weak Langmuir turbulence kinetic Eq.~(\ref{eq:KE_n_A}) (b).
Fiber losses (0.2~dB/km) have been included in (a)-(b).
(c)-(d) report the simulation (a)-(b) but in the absence of the fiber losses, so as to improve the visualization of the multimode collective behavior of the spectral oscillations of the discrete soliton.
Spectral profiles in normal scale (e), and logarithmic scale (f), at the propagation length $z=25$km corresponding to (c)-(d): the blue line reports the result of the NLS Eq.~(\ref{eq:A}) simulation (avergaed over the 6 fiber modes), the red line reports the result of the weak Langmuir turbulence kinetic Eq.~(\ref{eq:KE_n_A}) simulation, the dashed black line the initial condition.
Parameters are $\ell_c=2$m, $\sigma=3.1$m$^{-1}$, $P=8.5$W.
The good agreement between the simulations of the NLS Eq.~(\ref{eq:A}) and the kinetic Eq.~(\ref{eq:KE_n_A}) is obtained without adjustable parameters.
}
\label{fig:3} 
\end{figure}
\end{center}

\subsection{Synchronization of incoherent spectral oscillations}

To complete our study, we consider a more realistic numerical simulation in which the incoherent source launched into the MMF is characterized by a relatively narrow frequency bandwidth, $\Delta \nu \simeq 2$THz, which can be accessible from an amplified spontaneous emission (ASE) source, see e.g. \cite{PRA15}.
In addition, we have included the impact of the fiber losses in the numerical simulations, with a typical value of $0.2$dB/km.

The results of the numerical simulations of the generalized NLS Eq.(\ref{eq:nlsmulti}) and the kinetic Eq.(\ref{eq:KE_n_A}) are reported in Fig.~3(a)-(b).
As a consequence of the narrowness of the initial spectrum, the discrete nature of the spectral shift gets more apparent as compared to the discrete VSIS discussed above through Fig.~1.
This reinforces the idea of synchronization of the spectral oscillations of the fiber modes in the simulation of the NLS Eq.(\ref{eq:nlsmulti}).
Indeed, we have reported in Fig.~3(a) the spectrum averaged over the six fiber modes.
If the spectral oscillations were not synchronized, then the average among the six modes would be characterized by a significant spectral broadening.
The good agreement between the NLS and kinetic simulations shown in Fig.~3(a) and Fig.~3(b) reflects the accurate synchronisation among the spectral oscillations of the fiber modes.
Note that the fiber losses naturally induce a significant reduction of the nonlinear effects during the propagation, which thus limits the spectral red-shift of the field.
Accordingly, we have removed the fiber losses in the simulation reported in Fig.~3(c)-(d), which considerably improves the visualization of the synchronization of the spectral oscillations of the fiber modes.

We finally note a minor discrepancy between the simulation of the generalized NLS Eq.(\ref{eq:nlsmulti}) and the kinetic Eq.(\ref{eq:KE_n_A}) in Figs.~3(e)-(f).
This can be ascribed to the fact that the separation of scales between linear dispersion effects and nonlinear effects is only partially satisfied because of the narrowness of the spectrum considered in Fig.~3.
Then at variance with the simulation reported in Fig.~1 where $L_{lin} / L_{nl} \simeq 3.10^{-3}$, in the case of Fig.~3 we only have $L_{lin} / L_{nl} \simeq 0.1$, which can merely explain the slight discrepancy between the NLS and kinetic simulations observed in Figs.~3(e)-(f).

\section{Conclusion and discussion}

In summary, we have studied the propagation of spatio-temporal incoherent waves in MMFs in the presence of a random coupling among the modes.
By averaging over the fast disordered fluctuations, we have derived the multimode Manakov Eq.(\ref{eq:manakovB}-\ref{eq:ramanman2}).
The new Raman term (\ref{eq:ramanman2}) in the multimode Manakov equation unveils a coupling among the modes, which is responsible for the emergence of a collective multimode behavior of the incoherent waves.
Indeed, applying the wave turbulence theory to the multimode Manakov equation, we have derived a very simple scalar kinetic equation governing the evolution of the averaged spectrum of the multimode field.
The theory has been validated by the numerical simulations, which 
confirm the robustness of the process of modal attraction toward the dynamics described by the scalar kinetic equation: A quantitative agreement between the simulations of the NLS Eq.~(\ref{eq:A}) and the kinetic Eq.(\ref{eq:KE_n_A}) has been obtained, without using any adjustable parameters.
The simulations reveal that the fields that propagate in different modes of the MMF self-organize and self-trap to form a VSIS. 
In particular, the VSIS can exhibit a discrete behavior characterized by collective synchronized spectral oscillations in frequency space. 
This work should stimulate the realization of optical experiments in MMFs.
Aside from the discrete multimode spectral solitons, the reduction of the multimode NLS equation to the effective scalar kinetic Eq.(\ref{eq:KE_n_A}) can be exploited to study different turbulent regimes predicted in the scalar case, such as the formation of incoherent spectral shock waves \cite{PRL13}.

We recall that we have considered in this work the case of strong random coupling among the modes.
A similar analysis can be carried out by considering a weak random mode coupling, where only (quasi-)degenerate modes are coupled to each other.
Actually, weak random mode coupling is known to be relevant when relatively short propagation lengths in optical fibers are considered \cite{ho14}.
However, the validity of the kinetic approach requires a weak nonlinear regime, $L_{lin} \ll L_{nl}$, so that large propagation lengths, typically larger than a few hundred meters are required to observe the formation of multimode spectral incoherent solitons in optical fibers.
For such a large propagation length, it is commonly admitted that random coupling among non-degenerate modes should not be neglected and must be taken into account \cite{ho14}, which legitimizes the consideration of strong mode coupling in our work. 

We remark that the validity of the derived kinetic Eq.(\ref{eq:KE_n_A}) becomes questionable when the optical spectrum feels the presence of a zero-dispersion-frequency of the optical fiber.
Nearby a zero-dispersion-frequency, linear dispersive effects become perturbative.
The dynamics turns out to be dominated by nonlinear effects, which invalidates the weakly nonlinear assumption underlying the derivation of the kinetic equation.
In this case one needs to include higher-order contributions in the closure of the hierarchy of the moments equation in the wave turbulence theory. 
To next-order, the instantaneous Kerr nonlinearity coupled to higher-order dispersion effects leads to a collision term in the kinetic equation that describes an incoherent (turbulent) regime of supercontinuum generation \cite{PR14}.
It would be interesting to develop a generalized kinetic formulation of spatio-temporal effects in MMFs, which would unify the Langmuir formulation discussed here with the wave turbulence formulation accounting for random mode coupling discussed in \cite{PRL19,PRA19}.
Such a generalized theory can also shed new light on the recent experiments of supercontinuum generation that can be characterized by spatial beam cleaning effects \cite{krupa16,lopez_ol16,eftekhar17,niang19}.
From a broader perspective, this would contribute to the development of a wave turbulence theory that accounts for the presence of a structural disorder of the nonlinear medium \cite{PRL19,PRA19,cherroret15,nazarenko19,wang20}.




\section{Acknowledgements}

We acknowledge financial support from the French ANR under Grant No. ANR-19-CE46-0007 (project ICCI),
iXcore research foundation, EIPHI Graduate School (Contract No. ANR-17-EURE-0002), French program ``Investissement d'Avenir," Project No. ISITE-BFC-299 (ANR-15 IDEX-0003);  H2020 Marie Sklodowska-Curie Actions (MSCA-COFUND) (MULTIPLY Project No. 713694). Calculations were performed using HPC resources from DNUM CCUB (Centre de Calcul, Universit\'e de Bourgogne).


\bigskip

\appendix 

\begin{widetext}

\section{Derivation of the Manakov equation}

We derive the Manakov multimode NLS Eq.~(\ref{eq:manakovB}).
Without approximations, the vector field ${\bm B}(z,t)$  is solution of
\begin{eqnarray}
i \partial_z {\bm B} +{\bf U}^\dag {\bf D}^0  {\bf U} {\bm B} +i  {\bf U}^\dag {\bf V}  {\bf U} \partial_t {\bm B}
- {\bf U}^\dag {\bf W}  {\bf U}  \partial_t^2 {\bm B}
+  \gamma (1-f_R) {\bf U}^\dag {\bm P}({\bf U} {\bm B})
+  \gamma f_R  {\bf U}^\dag{\bm Q}({\bf U} {\bm B})  =0  ,
\end{eqnarray}
where
\begin{eqnarray*}
\big[ {\bf U}^\dag {\bf D}^0  {\bf U} {\bm B}\big]_p &=& \sum_{l} \Big[\sum_{p',n'} U^*_{p'p} D_{p'n'}^0 U_{n'l} \Big]  B_l , \\
\big[ {\bf U}^\dag {\bf V}  {\bf U} \partial_t  {\bm B}\big]_p &=& \sum_{l} \Big[\sum_{p',n'} U^*_{p'p} V_{p'n'} U_{n'l} \Big]  \partial_t B_l , \\
\big[ {\bf U}^\dag {\bf W}  {\bf U}  \partial_t^2 {\bm B}\big]_p &=& \sum_{l} \Big[\sum_{p',n'} U^*_{p'p} W_{p'n'} U_{n'l} \Big]  \partial_t^2 B_l , \\
\big[  {\bf U}^\dag {\bm P}({\bf U} {\bm B}) \big]_p &=& \sum_{l,m,n}  \Big[  \sum_{p',l',m',n'}
S_{p'l'm'n'}^K U_{p'p}^* U_{l'l} U_{m'm} U_{n'n}^* \Big] B_l B_m B_n^*  ,  \\
\big[ {\bf U}^\dag {\bm Q}({\bf U} {\bm B})  \big]_p &=& \sum_{l,m,n}  \Big[\sum_{p',l',m',n'}
S_{p'l'm'n'}^R U_{p'p}^* U_{l'l} U_{m'm} U_{n'n}^* \Big]   B_l [  R \star (  B_m B_n^*)]  .
\end{eqnarray*}
We assume here that the linear coupling between modes due to ${\bf D}(z)$ is strong enough so that this effect dominates and
the random matrix-valued process ${\bf U}(z)$ becomes uniformly distributed in the set of unitary matrices.
We can then replace the linear and nonlinear terms by the homogenized coefficients
\begin{eqnarray*}
\big[ {\bf U}^\dag {\bf D}^0  {\bf U} {\bm B}\big]_p &=& \sum_{l} \Big[\sum_{p',n'} \left<  U^*_{p'p}  U_{n'l} \right>  D_{p'n'}^0 \Big]  B_l , \\
\big[ {\bf U}^\dag {\bf V}  {\bf U} \partial_t  {\bm B}\big]_p &=& \sum_{l} \Big[\sum_{p',n'} \left<  U^*_{p'p}  U_{n'l} \right>  V_{p'n'} \Big]  \partial_t B_l , \\
\big[ {\bf U}^\dag {\bf W}  {\bf U}  \partial_t^2 {\bm B}\big]_p &=& \sum_{l} \Big[\sum_{p',n'} \left< U^*_{p'p}  U_{n'l} \right> W_{p'n'} \Big]  \partial_t^2 B_l , \\
\big[  {\bf U}^\dag {\bm P}({\bf U} {\bm B}) \big]_p  &=& \sum_{l,m,n}  \Big[  \sum_{p',l',m',n'}
S_{p'l'm'n'}^K \left< U_{p'p}^* U_{l'l} U_{m'm} U_{n'n}^* \right>\Big] B_l B_m B_n^*  ,  \\
\big[ {\bf U}^\dag {\bm Q}({\bf U} {\bm B})  \big]_p  &=& \sum_{l,m,n}  \Big[\sum_{p',l',m',n'}
S_{p'l'm'n'}^R\left<  U_{p'p}^* U_{l'l} U_{m'm} U_{n'n}^* \right> \Big]   B_l [ R \star ( B_m B_n^*)]  ,
\end{eqnarray*}
where the expectation is taken with respect to the stationary distribution of the random process ${\bf U}(z)$, that is 
the Haar measure on the unitary group in dimension $2N$.
Integration with respect to the Haar measure on the unitary group has been studied in the mathematical physics literature for a long time \cite{ullah,weingarten}.
A general formula for calculating monomial integrals is given in \cite{collins}.
In the case of monomials of rank 2 and 4, we have  \cite[Prop. 4.2.3]{hiai}:
\begin{eqnarray}
\left<  U_{ij} U_{i'j'}^* \right> &=& \frac{1}{2N} \delta_{ii'}\delta_{jj'} ,
\label{eq:uu}\\
\nonumber
\left<  U_{i_1j_1} U_{i_2j_2} U_{i_1'j_1'}^* U_{i_2'j_2'}^* \right>
&=& 
\frac{1}{4N^2-1} 
\big( 
\delta_{i_1i_1'} \delta_{i_2i_2'}\delta_{j_1j_1'}\delta_{j_2j_2'}
+
\delta_{i_1i_2'} \delta_{i_2i_1'}\delta_{j_1j_2'}\delta_{j_2j_1'}
\big)\\
&&
-\frac{1}{2N(4N^2-1)} 
\big( 
\delta_{i_1i_1'} \delta_{i_2i_2'}\delta_{j_1j_2'}\delta_{j_2j_1'}
+
\delta_{i_1i_2'} \delta_{i_2i_1'}\delta_{j_1j_1'}\delta_{j_2j_2'}
\big) .
\end{eqnarray}
Using these formulas we find
\begin{eqnarray*}
\big[ {\bf U}^\dag {\bf D}^0  {\bf U} {\bm B}\big]_p &=&  \widetilde{d}  B_p , \\
\big[ {\bf U}^\dag {\bf V}  {\bf U} \partial_t  {\bm B}\big]_p &=&  \frac{1}{\widetilde{v}}  \partial_t B_p , \\
\big[ {\bf U}^\dag {\bf W}  {\bf U}  \partial_t^2 {\bm B}\big]_p &=&  \frac{\widetilde{\beta}}{2} \partial_t^2 B_p , \\
\big[  {\bf U}^\dag {\bm P}({\bf U} {\bm B}) \big]_p &=& \sum_{l,m,n}  \widetilde{S}_{plmn}^K  B_l B_m B_n^* ,  \\
\big[ {\bf U}^\dag {\bm Q}({\bf U} {\bm B})  \big]_p &=& \sum_{l,m,n}   \widetilde{S}_{plmn}^R  B_l [ R \star( B_m B_n^*)]  ,
\end{eqnarray*}
with
\begin{eqnarray}
\widetilde{d} &=& \frac{1}{2N} {\rm Tr}({\bf D}^0),\\
\frac{1}{\widetilde{v}} &=& \frac{1}{2N} {\rm Tr}({\bf V}) ,\\  
\widetilde{\beta} &=& \frac{1}{N} {\rm Tr}({\bf W}) ,\\  
\nonumber
\widetilde{S}_{plmn}^{X} &=&\delta_{ln}\delta_{mp} \Big\{ \frac{1}{4N^2-1} \sum_{p',l'} S_{p'l'p'l'}^{X}- \frac{1}{2N(4N^2-1)}
\sum_{p',l'} S_{p'p'l'l'}^{X} \Big\} \\
&&+ \delta_{lp}\delta_{mn} \Big\{ \frac{1}{4N^2-1} \sum_{p',l'} S_{p'p'l'l'}^{X}- \frac{1}{2N(4N^2-1)}
\sum_{p',l'} S_{p'l'p'l'}^{X} \Big\} , \quad \quad X\in \{K,R\},
\nonumber
\end{eqnarray}
or equivalently
\begin{eqnarray*}
\widetilde{S}_{plmn}^{X} &=& \widetilde{S}^{X}_{(1)}  \delta_{ln}\delta_{mp}  
+   \widetilde{S}^{X}_{(2)} \delta_{lp}\delta_{mn} , \quad \quad X\in \{K,R\},
\end{eqnarray*}
with
\begin{eqnarray}
\widetilde{S}_{(1)}^{X} &=&    \frac{1}{4N^2-1} \sum_{p',l'} S_{p'l'p'l'}^{X}- \frac{1}{2N(4N^2-1)}
\sum_{p',l'} S_{p'p'l'l'}^{X}  , \quad \quad X\in \{K,R\},
\label{eq:Stilde1_KR} \\
\widetilde{S}_{(2)}^{X}  &=&\frac{1}{4N^2-1} \sum_{p',l'} S_{p'p'l'l'}^{X}- \frac{1}{2N(4N^2-1)}
\sum_{p',l'} S_{p'l'p'l'}^{X}  , \quad \quad X\in \{K,R\} .
\label{eq:Stilde2_KR}
\end{eqnarray}
In other words the mode amplitudes ${\bm B}$ satisfy
\begin{eqnarray}
i \partial_z {\bm B} +{\widetilde d} {\bm B} + \frac{i}{\widetilde{v}}  \partial_t {\bm B}
- \frac{\widetilde{\beta} }{2} \partial_{tt} {\bm B}
+  \gamma (1-f_R) \widetilde{\bm P}({\bm B})
+  \gamma f_R  \widetilde{\bm Q}({\bm B})  =0  ,
\end{eqnarray}
with
\begin{eqnarray}
\big[ \widetilde{\bm P}({\bm B}) \big]_p&=&
\big(  \widetilde{S}_{(1)}^K +  \widetilde{S}_{(2)}^K  \big)\Big[\sum_{l=1}^{2N} |B_l|^2 \Big]B_p
   ,\\
\big[ \widetilde{\bm Q}({\bm B}) \big]_p  &=&
 \widetilde{S}_{(1)}^R  \sum_{l=1}^{2N} B_l \big[ R \star( B_p B_l^*)\big] +
 \widetilde{S}_{(2)}^R  B_p \Big[ R \star \Big(\sum_{l=1}^{2N}  | B_l|^2\Big)\Big]  .
\end{eqnarray}

We remark that, if the initial conditions in  Eq.~(\ref{eq:manakovB}) are deterministic, then the components $|B_p|^2(z,t)$ are deterministic and  the components $A_p(z,t)$ of the field are random and given by ${\bm A}={\bf U} {\bm B}$, so that they verify $|A_p|^2(z,t)=\sum_{j,k}  U_{pj}(z) U^*_{pk}(z) B_j(z,t) B_k^*(z,t)$ and, using (\ref{eq:uu}), $\left<  |A_p|^2(z,t)\right> =\frac{1}{2N}\sum_{j} |B_j|^2(z,t)$  for any $t$ and $z$ much larger than the correlation length of ${\bf D}(z)$, say $\ell_c$.
In other words, we have equipartition in the ${\bm A}$ basis.
Note that this result is obtained for the {\it expectation} (with respect to the distribution  of $({\bf D}(z))_{z\geq 0}$) of the components $|A_p|^2(z,t)$.

We also remark that, if the initial conditions are random (and independent of $({\bf D}(z))_{z\geq 0}$), then the components $|B_p|^2(z,t)$ are governed by the deterministic Eq.~(\ref{eq:manakovB})  and  the components $A_p(z,t)$ are given by ${\bm A}={\bf U} {\bm B}$, so that they verify $|A_p|^2(z,t)=\sum_{j,k}  U_{pj}(z) U^*_{pk}(z) B_j(z,t) B_k^*(z,t)$ and, using (\ref{eq:uu}), $\left<  |A_p|^2(z,t)\right> =\frac{1}{2N}\sum_{j} \left<|B_j|^2(z,t)\right>$  for any $t$ and $z$ much larger than $\ell_c$.
Following the same remark, we have $\left< A_p(z,t+\tau/2) A_p^*(z,t-\tau/2)\right>=\frac{1}{2N}\sum_j \left<  B_j(z,t+\tau/2) B_j^*(z,t-\tau/2)\right>$  for any $p$ for $z \gg \ell_c$.

\end{widetext}




\begin{thebibliography}{99}

\bibitem{wright15np}
L. Wright, D. N. Christodoulides, and F.W. Wise, 
Controllable spatiotemporal nonlinear effects in multimode fibres, 
Nature Photon. 9, 306 (2015).


\bibitem{renninger13}
W. H. Renninger and F. W. Wise, 
Optical solitons in gradedindex multimode fibres, 
Nature Comm. 4, 1719 (2013).

\bibitem{krupa16_geom}
K. Krupa, A. Tonello, A. Barthelemy, V. Couderc, B. M. Shalaby, A. Bendahmane, G. Millot, and S. Wabnitz, 
Observation of geometric parametric instability induced by the periodic spatial self-imaging of multimode waves, 
Phys. Rev. Lett. {\bf 116}, 183901 (2016).

\bibitem{conforti17}
M. Conforti, C. Mas Arabi, A. Mussot, and A. Kudlinski,
Fast and accurate modeling of nonlinear pulse propagation in graded-index multimode fibers, 
Opt. Lett. 42, 4004 (2017).


\bibitem{krupa19}
K. Krupa {\it et al.,} 
Multimode nonlinear fiber optics, a spatiotemporal avenue, 
APL Photonics 4, 110901 (2019).

\bibitem{agrawal}
G. Agrawal, {\it Nonlinear Fiber Optics,} (Sixth Ed., Academic Press, New York, 2019).

\bibitem{kaminow13} 
I.P. Kaminow, T. Li, and A.F. Willner, 
{\it Optical Fiber Telecommunications, Systems and Networks} (Sixth Ed., Elsevier, 2013).

\bibitem{wright17sc}
L. G. Wright, D. N. Christodoulides, and F.W. Wise,
Spatiotemporal mode-locking in multimode fiber lasers,
Science 358, 94 (2017).

\bibitem{leonetti14}
M. Leonetti, S. Karbasi, A. Mafi, and C. Conti,
Light focusing in the Anderson regime,
Nature Comm. {\bf 5}, 4534 (2014).

\bibitem{schirmacher18}
W. Schirmacher, B. Abaie, A. Mafi, G. Ruocco, and M. Leonetti,
What is the right theory for Anderson localization of light? An experimental test,
Phys. Rev. Lett. {\bf 120}, 067401 (2018).

\bibitem{mafi19}
A. Mafi, J. Ballato, K.W. Koch, and A. Sch\"ulzgen,
Disordered Anderson Localization Optical Fibers for Image Transport - A Review,
J. Lightwave Techn. {\bf 37}, 5652-5659 (2019). 


\bibitem{silberberg08}
Y. Lahini, A. Avidan, F. Pozzi, M. Sorel, R. Morandotti, D.N. Christodoulides, and Y. Silberberg,
Anderson Localization and nonlinearity in one-dimensional disordered photonic lattices,
Phys. Rev. Lett. {\bf 100}, 013906 (2008).

\bibitem{segev}
M. Segev, Y. Silberberg, and D.N. Christodoulides, 
Anderson localization of light,
Nature Photonics {\bf 7}, 197 (2013).  


\bibitem{delre17}
D. Pierangeli, A. Tavani, F. Di Mei, A.J. Agranat, C. Conti, and E. DelRe,
Observation of replica symmetry breaking in disordered nonlinear wave propagation,
Nature Comm. {\bf 8}, 1501 (2017).


\bibitem{conti11}
C. Conti and L. Leuzzi,
Complexity of waves in nonlinear disordered media,
Phys. Rev. B {\bf 83}, 134204 (2011).

\bibitem{leuzzi15}
F. Antenucci, M. Ibanez Berganza, and L. Leuzzi,
Statistical physics of nonlinear wave interaction, 
Phys. Rev. B {\bf 92}, 014204 (2015).





\bibitem{psaltis} 
D. Psaltis and C. Moser, 
Imaging with multimode fibers,
Opt. Photonics News 27, 24 (2016).

\bibitem{wai96}
P.K.A. Wai and C.R. Menyuk,
Polarization mode dispersion, decorrelation, and diffusion in optical with randomly varying birefringence,
J. Lightwave Tech. {\bf 14}, 148-157 (1996).


\bibitem{mecozzi12a}
A. Mecozzi, C. Antonelli, and M. Shtaif, 
Nonlinear propagation in multimode fibers in the strong coupling regime, 
Optics Express {\bf 20}, 11673 (2012).

\bibitem{mecozzi12b}
A. Mecozzi, C. Antonelli, and M. Shtaif, 
Coupled Manakov equations in multimode fibers with strongly coupled groups of modes, 
Optics Express {\bf 20}, 23436 (2012).

\bibitem{mumtaz13}
S. Mumtaz, R.-J. Essiambre,  and G. P. Agrawal,
Nonlinear propagation in multimode and multicore fibers: Generalization of the Manakov equations,
J. Lightwave Tech. {\bf 31}, 398-406 (2013).

\bibitem{antonelli13}
C. Antonelli, A. Mecozzi, and M. Shtaif,
Raman amplification in multimode fibers with random mode coupling,
Opt. Lett. {\bf 38}, 1188 (2013).

\bibitem{buch19}
S. Buch, S. Mumtaz, R.-J. Essiambre, A. M. Tulino, and G. P. Agrawal, 
Averaged nonlinear equations for multimode fibers valid in all regimes of random linear coupling, 
Opt. Fiber Technol. {\bf 48}, 123 (2019).

\bibitem{antonelli16}
C. Antonelli, M. Shtaif, and A. Mecozzi,
Modeling of nonlinear propagation in space-division multiplexed fiber-optic transmission,
J. Lightwave Tech. {\bf 34}, 36-54 (2016).

\bibitem{liu16}
Z. Liu, L. G. Wright, D. N. Christodoulides, and F. W. Wise,
Kerr self-cleaning of femtosecond-pulsed beams in gradedindex
multimode fiber, Opt. Lett. {\bf 41}, 3675 (2016).

\bibitem{krupa17}
K. Krupa, A. Tonello, B. M. Shalaby, M. Fabert, A. Barth\'el\'emy, G. Millot, S. Wabnitz, and V. Couderc, 
Spatial beam self-cleaning in multimode fibres, 
Nat. Photonics {\bf 11}, 237 (2017).

\bibitem{terry07}
N.B. Terry, T.G. Alley, and T.H. Russell, 
An explanation of SRS beam cleanup in graded-index fibers and the absence of SRS beam cleanup in step-index fibers, 
Optics Express {\bf 15}, 17509 (2007).


\bibitem{PRL19}
A. Fusaro, J. Garnier, K. Krupa, G. Millot, and A. Picozzi,
Dramatic acceleration of wave condensation mediated by disorder in multimode fibers, 
Phys. Rev. Lett. {\bf 122}, 123902 (2019).

\bibitem{PRA19}
J. Garnier, A. Fusaro, K. Baudin, C. Michel, K. Krupa, G. Millot, and A. Picozzi, 
Wave condensation with disorder versus beam self-cleaning in multimode fibers,
Phys. Rev. A {\bf 100}, 053835 (2019).

\bibitem{pod19}
E. Podivilov, D. Kharenko, V. Gonta, K. Krupa, O. S. Sidelnikov, S. Turitsyn, M. P. Fedoruk, S. A. Babin, and S. Wabnitz, 
Hydrodynamic 2D Turbulence and Spatial Beam Condensation in Multimode Optical Fibers, 
Phys. Rev. Lett. {\bf 122}, 103902 (2019).

\bibitem{christodoulides19}
F. O. Wu, A. U. Hassan, and D. N. Christodoulides, 
Thermodynamic theory of highly multimoded nonlinear optical systems, 
Nat. Photon. {\bf 13}, 776 (2019).

\bibitem{kottos20}
A. Ramos, L. Fern\'andez-Alc\'azar, T. Kottos, and B. Shapiro,
Optical Phase Transitions in Photonic Networks: A Spin-System Formulation, 
Phys. Rev. X {\bf 10}, 031024 (2020).

\bibitem{PRL20}
K. Baudin, A. Fusaro, K. Krupa, J. Garnier, S. Rica, G. Millot, and A. Picozzi, 
Classical Rayleigh-Jeans condensation of light waves: Observation and thermodynamic characterization,
Phys. Rev. Lett. {\bf 125}, 244101 (2020).

\bibitem{EPL21}
K. Baudin, A. Fusaro, J. Garnier, N. Berti, K. Krupa, I. Carusotto, S. Rica, G. Millot, and A. Picozzi,
Energy and wave-action flows underlying Rayleigh-Jeans thermalization of optical waves propagating in a multimode fiber,
EPL {\bf 134}, 14001 (2021).

\bibitem{pourbeyram_arxiv}
H. Pourbeyram, P. Sidorenko, F. Wu, L. Wright, D. Christodoulides, and F. Wise, 
Direct measurement of thermalization to Rayleigh-Jeans distribution in optical beam self cleaning, ArXiv 2012.12110

\bibitem{zakharov92}
V.E. Zakharov, V.S. L'vov, G. Falkovich, 
{\it Kolmogorov Spectra of Turbulence I} (Springer, Berlin, 1992).


\bibitem{newell01} 
A.C. Newell, S. Nazarenko, L. Biven, 
Wave turbulence and intermittency, 
Physica D  {\bf 152}, 520 (2001).

\bibitem{nazarenko11} 
S. Nazarenko, {\it Wave Turbulence} (Springer, Lectures Notes in Physics, 2011).

\bibitem{Newell_Rumpf} 
A.C. Newell, B. Rumpf, 
Wave Turbulence,
Annu. Rev. Fluid Mech. {\bf 43}, 59 (2011).


\bibitem{shrira_nazarenko13}
{\it Advances in Wave Turbulence,} World Scientific Series on
Nonlinear Science Series A, Vol. 83, edited by V.I. Shrira and S. Nazarenko 
(World Scientific, Singapore, 2013).

\bibitem{PR14}
A. Picozzi, J. Garnier, T. Hansson, P. Suret, S. Randoux, G. Millot, and D. Christodoulides,
Optical wave turbulence: Toward a unified nonequilibrium thermodynamic formulation of statistical nonlinear optics,
Phys. Reports {\bf 542}, 1-132  (2014).

\bibitem{turitsyn12} 
E. Turitsyna, G. Falkovich, A. El-Taher, X. Shu, P. Harper, and
S. Turitsyn, Optical turbulence and spectral condensate in long
fibre lasers, Proc. R. Soc. London Ser. A {\bf 468}, 2145 (2012).

\bibitem{turitsyn15} 
D. Churkin, I. Kolokolov, E. Podivilov, I. Vatnik, S. Vergeles,
I. Terekhov, V. Lebedev, G. Falkovich, M. Nikulin, S. Babin, and S. Turitsyn, 
Wave kinetics of a random fibre laser, 
Nature Commun. {\bf 2}, 6214 (2015).



\bibitem{musher95}
S. Musher, A. Rubenchik, and V. Zakharov, 
Weak Langmuir turbulence, 
Phys. Rep. {\bf 252}, 177 (1995).

\bibitem{zakharovjetp}
V. E. Zakharov, 
Collapse of Langmuir waves,
Zh. Eksp. Teor. Fiz. {\bf 62}, 1745 (1972) 
[Sov. Phys. JETP {\bf 35}, 908 (1972)].

\bibitem{ZakhPR85}
V.E. Zakharov, S.L. Musher, A.M. Rubenchik,
Hamiltonian approach to the description of non-linear plasma phenomena,
Phys. Reports {\bf 129}, 285 (1985).



\bibitem{montes79}
C. Montes, Phys. Rev. A {\bf 20}, 1081 (1979).


\bibitem{PRL08}
A. Picozzi, S. Pitois, and G. Millot, 
Spectral incoherent solitons: A localized soliton behavior in frequency space, 
Phys. Rev. Lett. {\bf 101}, 093901 (2008).


\bibitem{OL16}
A. Fusaro, J. Garnier, C. Michel, G. Xu, J. Fatome, L. G. Wright, F. W. Wise, and A. Picozzi, 
Decoupled polarization dynamics of incoherent waves and bimodal spectral incoherent solitons, 
Opt. Lett. {\bf 41}, 3992 (2016).

\bibitem{PRE11}
B. Kibler, C. Michel, A. Kudlinski, B. Barviau, G. Millot, and A. Picozzi, 
Emergence of spectral incoherent solitons through supercontinuum generation in a photonic crystal fiber, 
Phys. Rev. E {\bf 84}, 066605 (2011).

\bibitem{GoldmanRMP}
M.V. Goldman,
Strong turbulence of plasma waves,
Rev. Mod. Phys. {\bf 56}, 709 (1984).

\bibitem{RobinsonRMP97}
P. A. Robinson, 
Nonlinear wave collapse and strong turbulence,
Rev. Mod. Phys. {\bf 69}, 507 (1997).

\bibitem{Langmuir38}
I. Langmuir,
Surface motion of water induced by wind.
Science {\bf 87}, 119 (1938).

\bibitem{Craik76}
A.D.D. Craik, S. Leibovich,
A rational model for Langmuir circulations,
J. Fluid Mech. {\bf 73}, 401 (1976).

\bibitem{Williams97}
J. C. Mc Williams, P. P. Sullivan, C.-H. Moeng,
Langmuir turbulence in the ocean,
J. Fluid Mech. {\bf 334}, 1 (1997).

\bibitem{Wong84}
A. Y. Wong, P. Y. Cheung, 
Three-Dimensional Self-Collapse of Langmuir Waves,
Phys. Rev. Lett. {\bf 52}, 1222 (1984).

\bibitem{Vyacheslavov02}
L. N. Vyacheslavov, V. S. Burmasov, I. V. Kandaurov, E. P. Kruglyakov, O. I. Meshkov, S. S. Popov, A. L. Sanin,
Strong Langmuir turbulence with and without collapse: experimental study,
Plasma Phys. Controlled Fusion {\bf 44}, B279 (2002).

\bibitem{Sulzer94}
M. P. Sulzer and J. A. Fejer,
Radar spectral observations of HF-induced ionospheric Langmuir turbulence with improved range and time resolution,
J. Geophys. Res. {\bf 99}, 15035 (1994).

\bibitem{Isham99}
B. Isham, C. La Hoz, M. T. Rietveld, T. Hagfors, and T. B. Leyser,
Cavitating Langmuir Turbulence Observed during High-Latitude Ionospheric Wave Interaction Experiments
Phys. Rev. Lett. {\bf 108}, 105003 (2012).


\bibitem{IshamPRL12}
B. Isham, M. T. Rietveld, P. Guio, F. R. E. Forme, T. Grydeland, E. Mjolhus,
Cavitating Langmuir Turbulence in the Terrestrial Aurora,
Phys. Rev. Lett. {\bf 108}, 105003 (2012).



\bibitem{horak}
P. Horak and F. Poletti,
Multimode nonlinear fibre optics: Theory and applications,
In Recent Progress in Optical Fiber Research, 
M. Yasin, S.W. Harun, H. Arof, eds., Intech, 3-25 (2012). 

\bibitem{ho14}
K.-P. Ho and J. M. Kahn, 
Linear propagation effects in mode-division multiplexing systems, 
J. Lightwave Tech. {\bf 32}, 4 (2014).


\bibitem{pourbeyram13}
H. Pourbeyram, G.P. Agrawal, and A. Mafi,
Stimulated Raman scattering cascade spanning the wavelength range of 523 to 1750nm using a graded-index multimode optical fiber,
Appl. Phys. Lett. {\bf 102}, 201107 (2013).


\bibitem{PRA11}
C. Michel, B. Kibler, and A. Picozzi, 
Discrete spectral incoherent solitons in nonlinear media with noninstantaneous response, 
Phys. Rev. A {\bf 83}, 023806 (2011).

\bibitem{Ramachandran19}
L. Rishoj, B. Tai, P.Kristensen, and S. Ramachandran,
Soliton self-mode conversion: revisiting Raman scattering of ultrashort pulses,
Optica {\bf 6}, 304 (2019).

\bibitem{agrawal19}
A. Antikainen, L. Rishoj, B. Tai, S. Ramachandran, and G.P. Agrawal, 
Fate of a soliton in a high order spatial mode of a multimode fiber,
Phys. Rev. Lett. {\bf 122} 023901 (2019).

\bibitem{OL14}
G. Xu, J. Garnier, M. Conforti, and A. Picozzi,
Generalized description of spectral incoherent solitons,
Opt. Lett. {\bf 39}, 4192 (2014).


\bibitem{PRA10}
J. Garnier and A. Picozzi, 
Unified kinetic formulation of incoherent waves propagating in nonlinear media with noninstantaneous response, 
Phys. Rev. A {\bf 81}, 033831 (2010).




\bibitem{PRA15}
M. Conforti, A. Mussot, J. Fatome, A. Picozzi, S. Pitois, C. Finot, M. Haelterman, B. Kibler, C. Michel, and G. Millot, 
Turbulent dynamics of an incoherently pumped passive optical fiber cavity: Quasisolitons, dispersive waves, and extreme events,
Phys. Rev. A {\bf 91}, 023823 (2015).

\bibitem{PRL13}
J. Garnier, G. Xu, S. Trillo, and A. Picozzi, 
Incoherent dispersive shocks in the spectral evolution of random waves, 
Phys. Rev. Lett. {\bf 111}, 113902 (2013).




\bibitem{krupa16}
K. Krupa, C. Louot, V. Couderc,M. Fabert, R. Guenard, B. M.
Shalaby, A. Tonello, D. Pagnoux, P. Leproux, A. Bendahmane,
R. Dupiol, G. Millot, and S. Wabnitz, 
Spatiotemporal characterization of supercontinuum extending from the visible to the mid-infrared in a multimode graded-index optical fiber, 
Opt. Lett. {\bf 41}, 5785 (2016).

\bibitem{lopez_ol16}
G. Lopez-Galmiche, Z. Sanjabi Eznaveh, M. A. Eftekhar, J. Antonio Lopez, L. G. Wright, F. Wise, D. Christodoulides, and R. Amezcua Correa,
Visible supercontinuum generation in a graded index multimode fiber pumped at 1064nm,
Opt. Lett. {\bf 41}, 2553-2556 (2016).


\bibitem{eftekhar17}
A. Eftekhar, L. G. Wright, M. S. Mills, M. Kolesik, R.
Amezcua Correa, F. W. Wise, and D. N. Christodoulides,
Versatile supercontinuum generation in parabolic multimode optical fibers, 
Optics Express {\bf 25}, 9078 (2017).

\bibitem{niang19}
A. Niang, T. Mansuryan, K. Krupa, A. Tonello, M. Fabert, P. Leproux, D. Modotto, O. N. Egorova, A. E. Levchenko, D. S. Lipatov, S. L. Semjonov, G. Millot, V. Couderc, and S. Wabnitz,
Spatial beam self-cleaning and supercontinuum generation with Yb-doped multimode graded-index fiber taper based on accelerating self-imaging and dissipative landscape,
Optics Express {\bf 27}, 24018 (2019).

\bibitem{cherroret15}
N. Cherroret, T. Karpiuk, B. Gr\'emaud, and C. Miniatura, 
Thermalization of matter waves in speckle potentials, 
Phys. Rev. A {\bf 92}, 063614 (2015).

\bibitem{nazarenko19}
S. Nazarenko, A. Soffer, and M.-B. Tran, 
On the wave turbulence theory for the nonlinear Schr\"odinger equation with random potentials, Entropy {\bf 21}, 823 (2019).

\bibitem{wang20}
Z. Wang, W. Fu, Y. Zhang, and H. Zhao,
Wave-turbulence origin of the instability of Anderson localization against many-body interactions, 
Phys. Rev. Lett. {\bf 124}, 186401 (2020).


\bibitem{ullah}
N. Ullah and C. Porter, 
Expectation value fluctuations in the unitary ensemble, 
Physical Review {\bf 132}, 948-950 (1963).


\bibitem{weingarten}
D. Weingarten, 
Asymptotic behavior of group integrals in the limit of infinite rank, 
Journal of Mathematical Physics {\bf 19}, 999-1001 (1978).

\bibitem{collins}
B. Collins and  P. Sniady, 
Integration with respect to the Haar measure on unitary, orthogonal and symplectic group, 
Commun. Math. Phys. {\bf 264}, 773-795 (2006). 

\bibitem{hiai}
F. Hiai and D. Petz, 
The semicircle law, free random variables and entropy, no. 77, American Mathematical Society, 2006.
















\end{thebibliography}

\end{document}